\documentclass[preprint,3p,authoryear,9pt]{elsarticle}

\usepackage{amssymb}

\journal{Journal of Theoretical Biology}

\usepackage{multicol,psfrag,xr,titleref}
\usepackage{graphicx,fancyhdr,subfigure}

\usepackage{rotating,theorem,amssymb,verbatim,amsmath}
\usepackage{fancyhdr,alltt,citesort}
\usepackage{array}

\newcommand{\one}{($i$) }
\newcommand{\two}{($ii$) }

\begin{document}

\begin{frontmatter}



\title{Negative Feedback and Physical Limits of Genes}


\author{Nicolae Radu Zabet$^{1,2,*}$ }

\address{$^1$School of Computing, University of Kent, CT2 7NF, Canterbury, UK\\
$^{2}$ Present address: Cambridge Systems Biology Centre and Department of Genetics, University of Cambridge, Tennis Court Road, Cambridge CB2 1QR, UK \\
$^*$ Corresponding author:  n.r.zabet@gen.cam.ac.uk
}

\begin{abstract}
This paper compares the auto-repressed gene to a simple one (a gene without auto-regulation) in terms of response time and output noise under the assumption of fixed metabolic cost. The analysis shows that, in the case of non-vanishing leak expression rate, the negative feedback reduces both the switching on and switching off times of a gene. The noise of the auto-repressed gene will be lower than the one of the simple gene only for low leak expression rates. Summing up, for low, but non-vanishing leak expression rates, the auto-repressed gene is both faster and less noisier compared to the simple one. 
\end{abstract}


\end{frontmatter}


\section{Introduction}

%
%
%
%
Genes which maintain a functional relationship between the concentration of the regulatory input protein(s) and the concentration of the output protein can be thought of as being capable of performing computations \citep{weiss_2003,buchler_2003,yokobayashi_2002,mayo_2006,fernando_2008}. 
For instance, often cells need to respond to the presence or the absence of various chemical factors. In this case, one can say that the cell performs some sort of "computations" on the input chemical factors and, based on the result of these computations, the cell will produce a specific chemical or physical response (output). 
A classic example is the lac operon in \emph{E.coli} where the proteins associated with the lactose metabolism are produced only when the glucose is absent and the lactose is present \citep{setty_2003}. This is often approximated by an AND NOT gate, which suggests that the operon performs logical computations.


These "computations" performed by genes are characterised by several properties (such as accuracy, speed and energy cost) which are significantly influenced by the specificity of the environment (the cell).  
For instance, in the case of low number of molecules, inherent fluctuations in reaction rates, caused by thermal noise, induce stochastic fluctuations (noise) in the copy number of molecules \citep{lei_2008}. Usually, in living cells, there are few copies of mRNA molecules, and one or two copies per gene \citep{arkin_1998}. Consequently, the gene expression process is affected by \emph{noise} \citep{kaern_2005}. In the context of genes as computational units, stochastic fluctuations can hide useful signals in noise and, thus, the accuracy of the response is reduced. 
%

In addition to accuracy, computations are also characterised by the speed at which they are performed \citep{bennett_1982}. 
The response of a gene to a change of input is not instantaneous, but rather affected by a time delay. This time delay is often called the \emph{response time} of the gene \citep{alon_book_2007} and is connected to the speed at which the genes "compute", in the sense that higher response times translate in slower computations, while lower response times in faster computations. 
In the case of genes, the speed at which a gene responds to changes in the transcription factors abundance is of high importance. In particular, a cell that is able to respond faster to changes in the environment can have certain advantages over slower cells. For example, a cell that is able to uptake food faster can  consume more nutrients compared to a slower cell and, consequently, have an energy advantage over it.   

Ideally, one would want to increase both speed and accuracy as much as possible, but this is often limited by the available energy supply \citep{bennett_1973,bennett_1982,lloyd_2000}. Each cellular process (protein production, protein decay and maintenance processes) has a \emph{metabolic cost} attached to it, which is, usually, measured in number of ATP molecules \citep{akashi_2002}. The notion of cost used in this paper is not the exact quantitative measure of the actual metabolic cost, but rather a number which describes how the actual metabolic cost scales when the parameters of the genes are changed. 

With few exceptions, these three properties (speed, accuracy and cost) were investigated previously in a stand-alone fashion. 
These exceptions include studies which analysed only the speed and accuracy and disregarded the cost in several molecular systems, such as: DNA based logic gates \citep{stojanovic_2003b}, protein-protein interaction networks \citep{wang_2010} and gene regulatory networks \citep{rosenfeld_2005,isaacs_2005,hooshangi_2005,shahrezaei_2008}. A few other studies examined all three properties (speed, accuracy and cost) in various gene regulatory networks, like:  auto-repressed genes \citep{stekel_2008}, toggle switches \citep{mehta_2008} or gene networks that used frequency encoded signals \citep{tan_2007b}. Nevertheless, these studies which integrate all three properties (speed, accuracy and cost) addressed different scenarios and used  a different measure of cost compared to the one used in this contribution (for a discussion on the measure of cost see below). For a comprehensive review on computational properties of molecular systems see \cite{zabet_2010_thesis}.

In addition to the aforementioned studies, \cite{zabet_2009} showed that, in the case of a simple gene (a gene without auto-regulation), the speed, accuracy and cost properties are interconnected. In particular, they found that under fixed metabolic cost there is a speed-accuracy trade-off, which is controlled by the decay rate, i.e., high decay rates lead to faster and less accurate responses while lower decay rates, to slower and more accurate ones. 

One of the central results of this previous work by \cite{zabet_2009} stated that the speed-accuracy trade-off is optimal for systems with zero leak rates, i.e., there are no solutions that have better speed-accuracy characteristics. 
The vanishing leak expression rate represents a theoretical performance limit of a gene under fixed cost, but it is difficult to achieve in real systems and would require high metabolic cost \citep{zabet_2009}. The current  paper will investigate whether the performance of a gene can be improved beyond this optimal configuration (of zero leak rate), i.e., if a gene can display faster response times and less noise at the output without increasing the metabolic cost. 

One candidate mechanism to enhance the performance of a gene is the negative feedback, which is a network motif in bacterial cells \citep{savageau_1974,thieffry_1998,shen_orr_2002,alon_book_2007}, in the sense that it is a sub-network which is encountered with high occurrence, e.g. in \emph{E.coli} $40\%$ of the genes are auto-repressed \citep{austin_2006}.  The fact that it is encountered with high occurrence suggests that auto-repressed genes have certain advantages compared to simple ones. This paper aims to investigate whether negative feedback can enhance both response time and output noise while keeping the metabolic cost fixed (equal to the one of the gene without auto-repression).

The auto-repressed gene is a well studied system, which received great attention from the community over the past decade \citep{alon_2007}. \cite{rosenfeld_2002} showed that negative feedback reduces only the response time of switching on (when the gene goes from a low expression to a high one) . In the current setting (genes as computational units), the switching direction is not important and, thus, the mechanism should be capable of reducing the response times of both switching on and switching off (when the gene goes from a high expression to a low one). One of the assumptions of the aforementioned study of \citep{rosenfeld_2002} is that genes do not display leaky expression, which is obviously not true for all genes. Actually, most of the genes will display non-zero leak expression rates, because the metabolic cost associated with removing the leak rate would be very high \citep{zabet_2009}. Thus, this paper aims to investigate whether, under the assumption of leaky expression, the negative feedback can reduce the response time of both switching on and off. 


%
Furthermore, experimental evidence suggested that a negatively auto-regulated gene displays lower noise compared with the gene without any type of auto-regulation \citep{becskei_2000}. Analytical results confirmed that the noise of the auto-repressed gene is lower than the one of the simple gene \citep{thattai_2001,paulsson_2004}. However, these two studies \citep{thattai_2001,paulsson_2004} did not consider fixed metabolic cost. Other studies derived the equation of noise analytically under the assumption that the simple gene and the auto-repressed one display equal average number of molecules at steady state \citep{paulsson_2000b,stekel_2008,zhang_2009b}. Their results confirmed that the auto-repressed gene reduces the output noise. Nevertheless, they assumed that if the two systems have an equal average number of molecules at steady state, they also have an equal metabolic cost. 

A better measure for the metabolic cost is the production rate of a gene \citep{zabet_2009,chu_2011}. 
This is justified by the fact that a measure of metabolic cost should describe the energy consumption per time unit. For example, consider the case of two proteins ($X1$ and $X2$) that have the same average number of molecules at steady state, but the first one ($X1$) is produced and decayed faster compared with the second one ($X2$). Then, more molecules of the first protein ($X1$) will be produced and decayed compared with the second one ($X2$) and, consequently, the metabolic cost associated with the first protein ($X1$)  will be higher compared with the one of the second protein ($X2$). Hence, the production rate will describe better the scaling properties of the metabolic cost compared to the average number of molecules at steady state.  



This paper aims to compare systems that have equal metabolic cost. In the case of fixed decay rate, imposing the production rates (the measure of metabolic cost) of two systems to be equal leads to the output steady state abundances of the systems to be equal as well. This means that, when production rate is kept fixed, it is ensured that a previously used measure of cost (steady state abundance) is also kept fixed. Nevertheless, the current analysis is not limited to cases where the decay rate is kept fixed (see for example Figure \ref{fig:speedAccuracyInstant})  and, in those cases,  attempting to keep production rates fixed will lead to variable steady state abundances. Due to the reasons mentioned above, the production rate will be used as the only indicator of metabolic cost when the decay rate is not fixed. Note however that, for fixed decay rate, the two measures are equivalent and, consequently, the results obtained under the assumption of fixed production rates (metabolic cost) are also valid for equal steady state abundances.

The results from this contribution show that negative feedback reduces the response time in the case of leaky expression for both switching on and switching off. In addition, for low leak rates, negative feedback reduces the noise, while for high leak rates it increases the noise. Both these results were obtained under the assumption of fixed metabolic cost. 
Furthermore, the analysis identified a subset in the parameter space (low but non-vanishing leak rates) where the negatively auto-regulated gene outperforms the simple one in both speed and accuracy, thus, setting a new theoretical performance limit for genes.

\section{Model}


\subsection{Simple Gene}
\begin{figure}[htp]
  \begin{center}
	\subfigure[simple gene]{\includegraphics[width=0.23\textwidth]{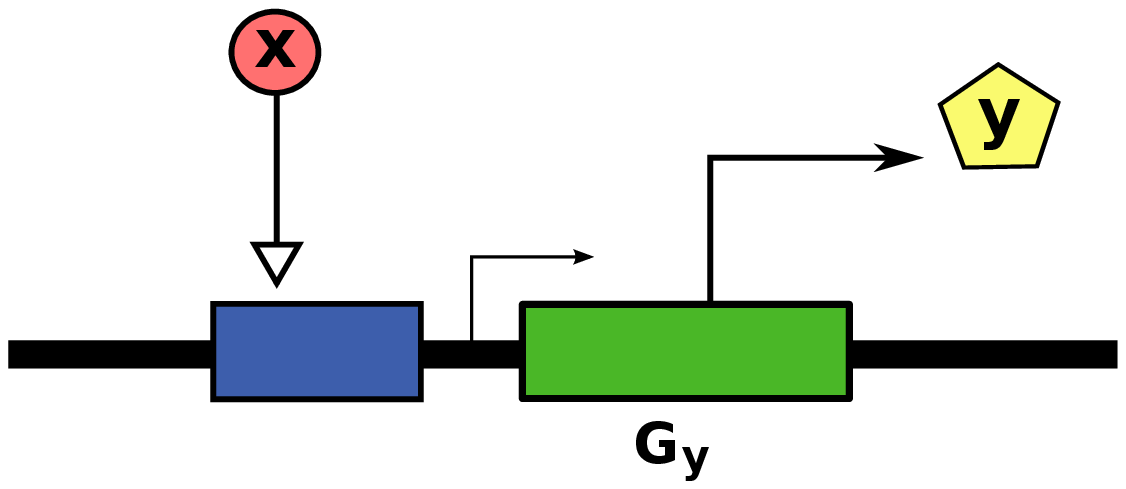}\label{fig:model-simple}}
	\subfigure[auto-repressed gene]{\includegraphics[width=0.23\textwidth]{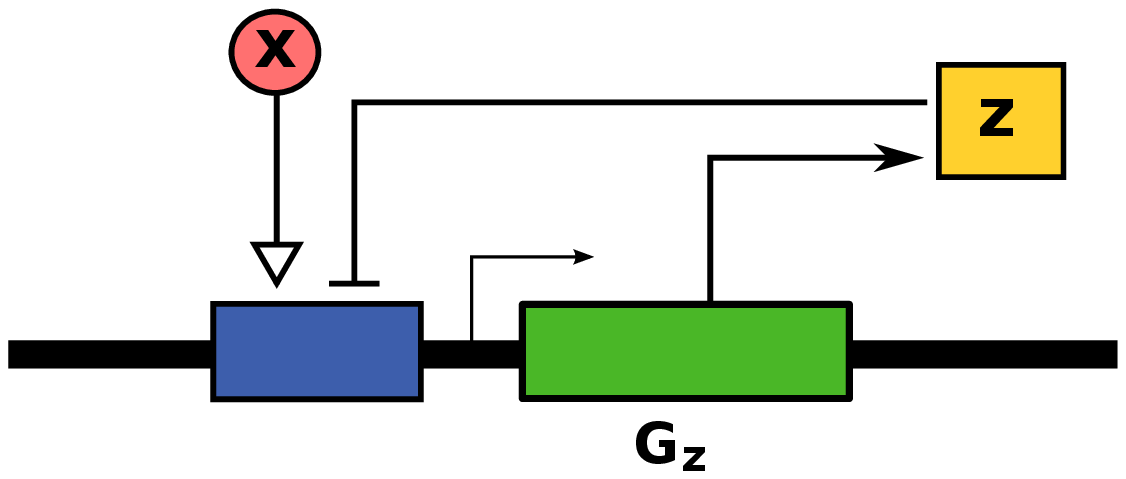}\label{fig:model-nar}}
  \end{center}
\caption[The model]{\emph{The model}. \subref{fig:model-simple} Gene $G_y$ activity is regulated only by protein $x$. \subref{fig:model-nar} The output protein of an auto-repressed gene is regulated by protein $x$ and down regulated by its own product protein.}\label{fig:model}
\end{figure}

The model of the simple gene consists of a single gene $G_y$,  which has an output  $y$ and is regulated by a single input species $x$; see Figure \ref{fig:model-simple}. This system is described by the following set of chemical reactions
\begin{equation}
\emptyset \xrightarrow{\beta f(x)}y, \quad y \xrightarrow{\mu}\emptyset\label{eq:modelSimpleReact}
\end{equation}
Here, $\beta$ is the maximal expression rate of the gene, $f(x)$ the regulation function, $x$ the concentration of the regulatory input, and $\mu$ is the degradation rate of the product of the gene; see Table \ref{tab:nomenclature}. 

Gene regulation functions are often approximated by Hill functions \citep{ackers_1982,bintu_2005_model,chu_2009}, which are sigmoid functions characterized by two parameters, namely the threshold ($K$) and the Hill coefficient ($l$). The latter determines the steepness of the function, whereas $K$ represents the required input concentration to achieve half activation of the gene.  This paper considers two  families of regulation functions, namely  the Hill activation function ($f\equiv\phi$) and the Hill repression function ($f\equiv\bar{\phi}$).
\begin{equation}
\label{eq:modelHill}
{\phi}(x) = \frac{x^l}{K^l  + x^l}\qquad\textrm{and}\qquad \bar{\phi}(x) = \frac{K^l}{K^l  + x^l}
\end{equation}

\begin{table*}[h]
 \begin{center}
  \begin{tabular}{|>{\centering}m{1.7cm}| m{11.8cm} |}
        \hline       
        $x$        								     									&  abundance of the regulatory input protein \tabularnewline[1pt]\hline
        $y$         								    									&  output abundance of the simple gene \tabularnewline[1pt]\hline
        $z$           								  									&  output abundance of the auto-repressed gene \tabularnewline[1pt]\hline
        $\beta$     								 									&  maximum expression rate for both genes \tabularnewline[1pt]\hline
        $\mu$  									     									&  decay rate of the output proteins \tabularnewline[1pt]\hline
        $f(x)$   												     						&  gene activation function \tabularnewline[1pt]\hline
        $K$           								 									&  threshold of the gene activation function \tabularnewline[1pt]\hline
        $l$             																	&  Hill coefficient of the gene activation function \tabularnewline[1pt]\hline
        $g(z)$  									     		 							&  gene auto-repression function \tabularnewline[1pt]\hline
        $K_n$   										     							&  threshold of the gene auto-repression function \tabularnewline[1pt]\hline
        $x_L$     								    									&  abundance level of the input that leads to low gene expression\tabularnewline[1pt]\hline
        $x_H$       								 									&  abundance level of the input that leads to high gene expression \tabularnewline[1pt]\hline
        $H$          									  									&  high abundance level of the output for both the simple and auto-repressed genes \tabularnewline[1pt]\hline
        $L$           																  	&  low abundance level of the output for the simple gene\tabularnewline[1pt]\hline
        $L_n$       											  							&  low abundance level of the output for the auto-repressed gene\tabularnewline[1pt]\hline
        $m$        		    															&  relative leak rate, $m=L/H$ or $m=L_n/H$\tabularnewline[1pt]\hline
        $x_0,\ y_0,\ z_0$															&  initial steady state abundance of the specific protein\tabularnewline[1pt]\hline
        $x^*,\ y^*,\ z^* $														&  new steady state abundance of the specific protein\tabularnewline[1pt]\hline
    	$y_\theta,\ z_\theta$														&  fraction $\theta$ of the new steady state abundance\tabularnewline[1pt]\hline
        $\tilde{t},\ \tilde{\beta},\ \tilde{T}$								&  specific parameters when the time is measured in $1/\mu$ (change of variable)\tabularnewline[1pt]\hline
        $T_y^{LH},\ T_z^{LH}$												&  switching on time\tabularnewline[1pt]\hline
        $\ T_y^{HL},\ T_z^{HL}$ 									 			&  switching off time\tabularnewline[1pt]\hline
        $T_d^{LH},\ T_d^{HL}$		 										&  difference between the switching time of the simple gene and the one of the auto-repressed gene \tabularnewline[1pt]\hline
        $\sigma^2_{y},\ \sigma^2_{z}$									&  variance of the steady state abundance \tabularnewline[1pt]\hline
        $\eta_{y},\ \eta_{z}$													&  normalized variance (noise) of the steady state abundance\tabularnewline[1pt]\hline
        $\eta^{\textrm{in}}_{y},\ \eta^{\textrm{in}}_{z}$		&  intrinsic component of noise \tabularnewline[1pt]\hline
        $\eta^{\textrm{up}}_{y},\ \eta^{\textrm{up}}_{z}$ 	&  upstream (from input) component of noise\tabularnewline[1pt]\hline
        $\eta_{c}^{\textrm{in}},\ \eta_{c}^{\textrm{up}}$		&  ratio between the noise of the auto-repressed gene and the one of the simple gene\tabularnewline[1pt]\hline
  \end{tabular}
 \end{center}
\caption{Nomenclature} \label{tab:nomenclature}
\end{table*}

\begin{figure*}[htp]
  \begin{center}
	\subfigure[steady state]{\includegraphics[scale=1.0]{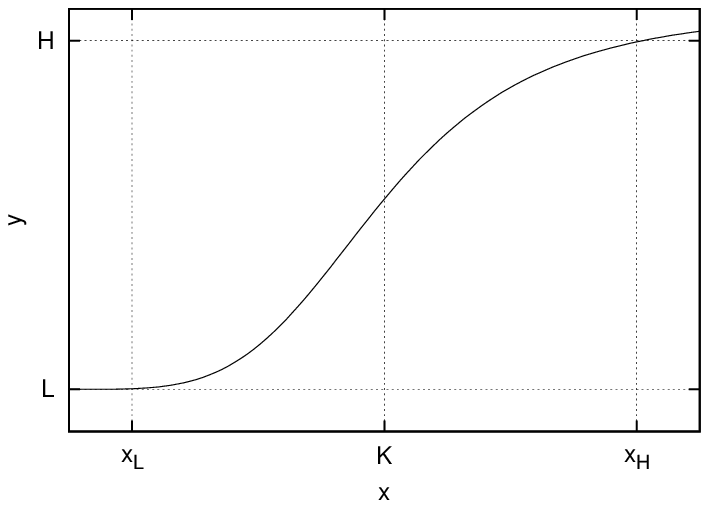}\label{fig:modelSimple-ss}}
	\subfigure[dynamic behaviour]{\includegraphics[scale=1.0]{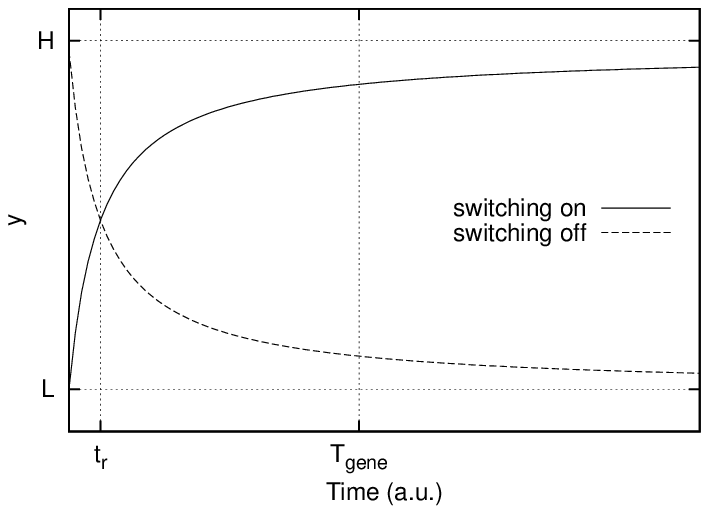}\label{fig:modelSimple-dynamic}}
  \end{center}
\caption[Steady state and dynamic behaviour of the simple gene]{\emph{Steady state and dynamic behaviour of the simple gene}. \subref{fig:modelSimple-ss} The output concentration of gene $G_y$ depends on the abundance of the regulatory protein $x$.  The activation case was considered.  \subref{fig:modelSimple-dynamic} The input of the system ($x$) changes instantaneously between a low state $x_L$ (corresponding $y = L$) and a high state $x_H$ (corresponding to $y = H$). In the case of switching on $y$ evolves from $y_0=L$ to $y^*=H$, while in the case of switching off $y$ evolves from $y_0=H$ to $y^*=L$. $t_r$ is the response time, which was defined by \cite{rosenfeld_2002} as the time to reach $50\%$ activation. $T_{gene}$ is the time to reach a fraction ($90\%$ in this graph) of the new steady state \citep{zabet_2009}.}\label{fig:modelSimple}
\end{figure*}

Usually, the equilibrium state (steady-state) and the kinetic behaviour (time evolution) of a system are defined using the differential equation associated to the system (the ODE) \citep{b_murray_2002}. In the case of the simple gene, the ODE has the following form
\begin{equation}
\label{eq:modelSimpleDiff}
\frac{d y(t)}{dt} = \beta f(x) - \mu y(t) 
\end{equation}
This system is completely determined by the input $x$, which is assumed to change instantaneously between two abundance levels, $x=x_L$ and $x=x_H$ \citep{zabet_2009,chu_2011}. These two levels of abundances of $x$ will lead to two concentration levels of the output $y$, namely a high concentration state ($y=H$ when $x=x_H$) and a low one ($y=L$ when $x=x_L$); see Figure \ref{fig:modelSimple-ss}.
For convenience, the low state (also called the leak rate) will be denoted as the fraction $m$ from the high state
\begin{equation}
L = m H, \quad m\in[0,1] \label{eq:modelSimpleSSL}
\end{equation} 
where by restricting $m$ to the interval $[0,1]$ it is ensured that $L\leq H$.

Note that \cite{rosenfeld_2002} assumed that the concentration of the output evolved from an initial value of zero ($y_0=L=0$) to a new non-zero one ($y^{*}=H>0$), which led to the following limitations: \one they only consider the switching on (or raise) time and \two they assume that there is no leaky gene expression. The current paper assumes a more general scenario, in which both the initial ($y_0$) and the new steady state ($y^{*}$) have positive values ($y_0 \geq 0$ and $y^{*} \geq 0$) and the system can evolve in both directions, i.e., in the case of switching on, the gene evolves from a low output state ($y_0=L$) to high one ($y^*=H$), while, in the case of switching off, the gene evolves from a high output state ($y_0=H$) to a low one ($y^*=L$); see Figure  \ref{fig:modelSimple-dynamic}.

\subsection{Auto-Repressed Gene}
A negatively auto-regulated gene is a gene which synthesises a protein that represses its own synthesis; see Figure \ref{fig:model-nar}. The auto-repressed gene is denoted by $G_z$ and this gene synthesises protein $z$. The differential equation that describes this system becomes
\begin{equation}
\frac{dz}{dt}=\beta f(x)g(z) -\mu z \label{eq:modelNarDiff}
\end{equation}
where $g(z)$ is a Hill repression function,
\begin{equation*}
g(z) = \gamma\frac{K_n^{h}}{K_n^{h}+z^{h}}.
\end{equation*}
The auto-repression function, $g(z)$, adds three new parameters, namely: $\gamma$ the maximum contribution of the auto-repression function ($\beta\gamma$ is the maximum production rate), $K_n$ the auto-repression threshold and $h$ the auto-repression Hill coefficient. In order to keep the mathematics trackable, this function needs to be brought to a simpler form. 
First, similarly as in the paper of \cite{rosenfeld_2002}, it is assumed that the auto-repression Hill coefficient is $1$. This yields 
\begin{equation}
g(z) = \gamma\frac{K_n}{K_n+z}
\end{equation}
Note that bacterial cells can implement auto-repression when the output protein binds to the promoter area of the gene and stops the RNAp molecules to transcribe the gene. In this case and assuming that only monomers auto-regulate the gene, the Hill coefficient of the auto-repression function is $1$ \citep{chu_2009}.

Furthermore, to make the comparison "fair", the two systems (the simple gene and the negatively auto-regulated one) will display equal metabolic costs.
As previously \citep{zabet_2009,chu_2011}, it is assumed that the \emph{metabolic cost} of a gene is approximated by the maximum production rate, which represents the pessimistic scenario. This maximal production rate is achieved when $x=x_H$, which results in the abundance of the output proteins reaching the maximum value of $y=z=H$. 
%
%
%
Note that the exact metabolic cost will depend on various additional factors, such as the length of the encoded protein, the average time the gene is active and so on. Nevertheless, these details are irrelevant in the current scenario, where only systems with single genes are considered.

Since the metabolic cost ($\zeta$) is measured as the maximum production rate (when $x=x_H$), imposing that the two systems have equal metabolic costs yields
\begin{equation*}
\zeta = \beta \cdot f(x_H)\cdot g(H) = \beta \cdot f(x_H)
\end{equation*}
which leads to
\begin{equation}
g(z) = \frac{K_n + H}{K_n + z} \label{eq:modelNarG}
\end{equation}
where $\gamma$ was set to $\gamma = (K_n+H)/K_n$ in order to ensure that the two systems have equal metabolic cost. Hence, the new parameter $\gamma$ was removed from the model.

The steady state abundance levels of the output $y$ which corresponds to the input concentration levels of $x=x_L$ and $x=x_H$ are computed as follows. First, for $x=x_H$, it results from equation \eqref{eq:modelNarG} that $g(z)=1$ and, thus, the differential equation \eqref{eq:modelNarDiff} becomes
\begin{equation*}
\frac{dz}{dt}=\beta f(x) -\mu z 
\end{equation*}
This yields
\begin{equation}
H_n = \frac{\beta}{\mu}f(x_H)=H\label{eq:modelNarSS-H}
\end{equation}
, which led to the same steady state abundance as in the case of the simple gene.

Furthermore, feeding equation \eqref{eq:modelNarG} into the differential equation \eqref{eq:modelNarDiff} and solving at steady state for $x=x_L$ leads to only one positive solution: 
\begin{equation}
L_n = \frac{1}{2}\left(- K_n + \sqrt{ K_n^2 + 4\frac{\beta}{\mu}f(x_L)\left(K_n + H\right)}\right) \label{eq:modelNarSS-L}
\end{equation}
The low output state of the negatively auto-regulated gene is denoted by $L_n$, as opposed to $L$, the low output state of the simple gene. Since both systems display the same abundance in the high output state, this high output state will be denoted by $H$.

From the form of equations \eqref{eq:modelNarSS-H} and \eqref{eq:modelNarSS-L} one can see that the high output state remains constant while changing the strength of the auto-repression, $K_n$, but the low state is increased if the auto-repression is strengthened ($K_n \searrow\ \Rightarrow\ L_n \nearrow$). It can be shown that the low state of the output varies between the following limits
\begin{eqnarray}
\lim_{K_n\rightarrow \infty} L_n & = &  \frac{\beta}{\mu}f(x_L) = L\nonumber\\ 
\lim_{K_n\rightarrow 0} L_n & = & \frac{\beta}{\mu}\sqrt{f(x_L)f(x_H)} = \sqrt{L \cdot H}=\sqrt{m} \cdot H 
\end{eqnarray}

Depending on the relationship between $K_n$ and $H$, the auto-regulation function \eqref{eq:modelNarG} can be rewritten in a simpler form. In particular, there are two extreme cases: \one $K_n\gg H$ and \two $K_n\ll H$. In the limit of weak auto-repression ($K_n\gg H$), the system will be similar to the simple one. This case does not pose any interest, since this paper aims to compare the simple gene to a system that displays different behaviour.

Furthermore, for $K_n \ll H$, the auto-repression becomes strong and the auto-regulation function is approximated by
\begin{equation}
g(z) = \frac{K_n + H}{K_n + z} \approx \frac{H}{z} \label{eq:modelNarGStrong}
\end{equation}
Note that the current definition of strong auto-repression is slightly different from the one of \cite{stekel_2008}, which considers strong auto-repression in absolute values, i.e., smaller than $< 10^{-4}\ \mu M$. The current definition is rather concerned with the relative repression strength ($K_n$) compared with the high abundance level of the output species ($H$) and aims to determine a parameter space where the auto-regulation function \eqref{eq:modelNarG} can be written in a simpler form \eqref{eq:modelNarGStrong}.

In this case, $K_n \ll H$, the low output state will be approximated 
\begin{equation}
L_n \approx \sqrt{m} H \label{eq:modelNarSSL}
\end{equation}

The auto-repression function specified in equation \eqref{eq:modelNarGStrong} does not contain any of the three additional parameters ($\gamma$, $K_n$ and $h$) and, thus, it represents a simpler mathematical formulation of the system. This form of the auto-repression function will be used, in the next section,  to compare the two systems (simple gene and auto-repressed one) in both speed and accuracy.

\section{Response Time}
Generally, one would want to process information as fast as possible, but genes are very slow, in the sense that the time required to turn on/off a gene (the switching time) is of the order of tens of minutes, even for an instant input change. Thus, it is essential to investigate what constrains the speed at which genes function and whether there are any methods to increase this speed. 

A common measure of the processing speed of genes is the \emph{response time}; that is the time required for the output of a gene to reach a new steady state once the input was changed. Note that the regulatory input, $x$, evolves from the initial abundance state $x_0$ to the new one $x^*$ (switching between $x_L$ and $x_H$). 
The input $x$ can either change instantaneously or non-instantaneously and, below, both cases will be considered.

\subsection{Instantaneous Change of Input}
The time of the simple gene to reach a fraction $\theta$ of the steady state when the input is changed instantaneously was computed previously by \cite{zabet_2009} as
\begin{equation}
T_y=\frac{1}{\mu}\ln\frac{y^*-y_0}{y^*-y_\theta}
\end{equation}
where $y_\theta = y_0+(y^*-y_0)\theta$.

In the case of the auto-repressed gene, knowing that the auto-repression function is the one from equation \eqref{eq:modelNarGStrong}, the solution to the differential equation \eqref{eq:modelNarDiff} yields
\begin{equation}
z(t) = \sqrt{\frac{\beta}{\mu}f(x)H + \left[(z_0)^2-\frac{\beta}{\mu}f(x)H\right]e^{-2\mu t}}
\end{equation}
The time to reach a fraction $\theta$ of the steady state, $z_\theta = z_0+(z^*-z_0)\theta$,  becomes
\begin{equation}
T_z=\frac{1}{2\mu}\ln\frac{\frac{\beta}{\mu}f(x)H-\mu^2 (z_0)^2}{\frac{\beta}{\mu}f(x)H-\mu^2 (z_\theta)^2} = \frac{1}{2\mu}\ln\frac{(z^*)^2-\mu^2 (z_0)^2}{(z^*)^2-\mu^2 (z_\theta)^2}
\end{equation}
where the steady state solution for the output protein $z$ was used, $z^* = \sqrt{\beta f(x) H/\mu}$.

To make the comparison easier, a change of variable will be performed to the differential equations attached to the two systems, \eqref{eq:modelSimpleDiff} and \eqref{eq:modelNarDiff}. The time $t$ will be measured in units of $1/\mu$ and this new time is denoted by $\tilde{t} = t/\mu$. The two differential equations become
\begin{eqnarray}
\frac{dy}{d \tilde{t}} &=& \tilde{\beta} f(x) -y\\
\frac{dz}{d \tilde{t}} &=& \tilde{\beta} f(x)\frac{H}{z} -z
\end{eqnarray}
Recomputing the time to reach a fraction $\theta$ of the steady state yields
\begin{eqnarray}
\tilde{T}_y &=& \ln\frac{y_0 - y^*}{y_\theta-y^*} \label{eq:timeReduced-simple}\\
\tilde{T}_z &=& \frac{1}{2}\ln\frac{(z^*)^2- (z_0)^2}{(z^*)^2-(z_\theta)^2}\label{eq:timeReduced-nar}
\end{eqnarray}

In the case of the simple gene, the switching on and switching off are equal. However, for the auto-repressed gene, these two response times are usually different. The response time of the gene is measured as the maximum between switching on time  and switching off time \citep{zabet_2010b}.

\subsubsection{Switching On}
First, the case when the two systems are turned on will be considered, i.e., ($y_0 = L$, $y^* = H$) and ($z_0 = L_n$, $z^* = H$). The fraction $\theta$ of the steady state can be written as:
\begin{eqnarray}
y_\theta&=&L+(H-L)\theta = H\left[m+(1-m)\theta\right]\nonumber\\
z_\theta&=&L_n+(H-L_n)\theta = H\left[\sqrt{m}+(1-\sqrt{m})\theta\right]
\end{eqnarray}
where the following equations where used $L=m\cdot H$ and $L_n=\sqrt{m}\cdot H$

Using equations \eqref{eq:timeReduced-simple} and \eqref{eq:timeReduced-nar} the switching on times yield
\begin{eqnarray}
\tilde{T}_y^{LH} &=& \ln{\frac{1}{1-\theta}}\nonumber\\
\tilde{T}_z^{LH} &=& \frac{1}{2}\ln{\frac{1-m}{1-\left[\sqrt{m}+(1-\sqrt{m})\theta\right]^2}}
\end{eqnarray}

\begin{figure*}[htp]
\psfrag{timedifference}{$\tilde{T}_d^{LH}$}
  \begin{center}
	\subfigure[switching on]{\includegraphics[scale=0.82]{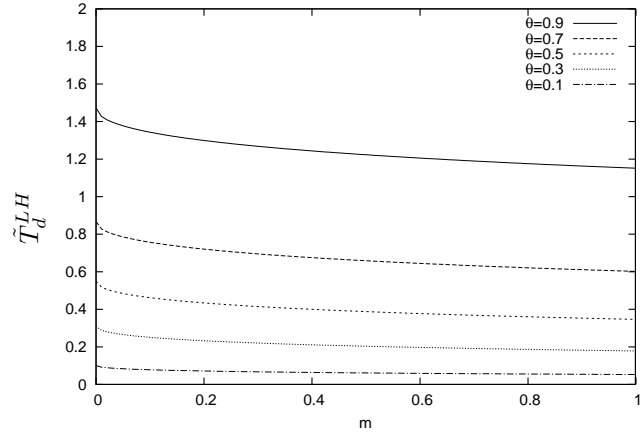}\label{fig:timeNarSimpleDifference-on}}
	\subfigure[switching off]{\includegraphics[scale=0.82]{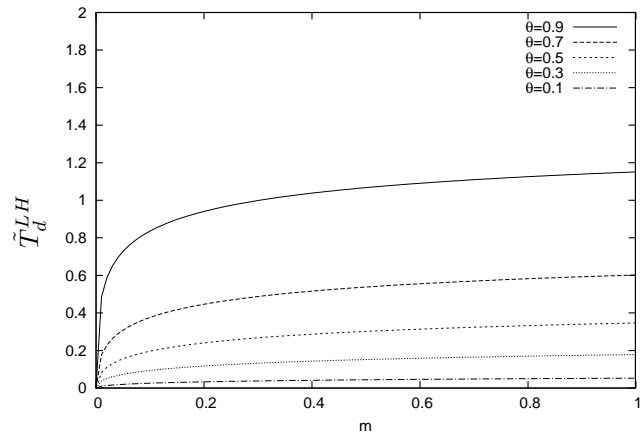}\label{fig:timeNarSimpleDifference-off}}
  \end{center}
\caption[Negative auto-regulation enhances the switching speed]{\emph{Negative auto-regulation enhances the switching speed}. For any combination $(\theta, m) \in[0,1]^2$ the difference between the switching time of the simple gene and of the auto-repressed gene is positive. This means that  the auto-repressed gene is faster than the simple gene. Note that this time difference is measured in $1/\mu$ and, thus, the actual time enhancement scales by $1/\mu$, i.e., $T_d^{LH}=\tilde{T}_d^{LH}/\mu$.}\label{fig:timeNarSimpleDifference}
\end{figure*}

The two times to reach a fraction of the steady state are compared by computing the difference between the two
\begin{eqnarray*}
\tilde{T}_d^{LH} &=& \tilde{T}_y^{LH} - \tilde{T}_z^{LH} \nonumber\\
 &=& \frac{1}{2}\ln{\left[\frac{1}{(1-\theta)^2} \frac{1-\left[\sqrt{m}+(1-\sqrt{m})\theta\right]^2}{1-m}\right]}\nonumber\\
 &=& \frac{1}{2}\ln{\frac{1-m-2\sqrt{m}\theta+2m\theta-\theta^2+2\sqrt{m}\theta^2-m\theta^2}{1-m-2\theta+2m\theta+\theta^2-m\theta^2}}
\end{eqnarray*}
This time difference is positive (and consequently negative auto-regulation speeds up the switching on) when the fraction in the logarithm is higher than or equal to $1$, which takes place when
\begin{equation*}
\theta(1-\theta)(1-\sqrt{m})\geq 0
\end{equation*}
This is true for any $\theta, m \in[0,1]$. Hence, negative auto-regulation always speeds up the switching on time compared with the simple gene. Figure \ref{fig:timeNarSimpleDifference-on} confirms  this result and shows that higher fractions $\theta$ of the steady states display better increase in speed that lower ones. 

In the special case of no leak rate (the optimum configuration for noise), $L=0$ and $L_n=0$,  the time gain reduces to
\begin{equation}
\tilde{T}_d^{LH} = \frac{1}{2}\ln{\frac{1+\theta}{1-\theta}}
\end{equation}
which is positive as long as $\theta >0$.

\subsubsection{Switching Off}
When the gene is turned off, ($y_0=H$, $y^*=L$) and ($z_0=H$, $z^*=L_n$), the output states ($H$, $L$ and $L_n$) remain the same as the ones for switching on, but the fractions $\theta$  of the steady state become
\begin{eqnarray}
y_\theta&=&H+(L-H)\theta = H\left[1-(1-m)\theta\right]\nonumber\\
z_\theta&=&H+(L_n-H)\theta = H\left[1-(1-\sqrt{m})\theta\right]
\end{eqnarray}
From equations  \eqref{eq:timeReduced-simple} and \eqref{eq:timeReduced-nar} one can compute the switching off time as
\begin{eqnarray}
\tilde{T}_y^{HL} &=& \ln{\frac{1}{1-\theta}}\nonumber\\
\tilde{T}_z^{HL} &=& \frac{1}{2}\ln{\frac{1-m}{\left[1-(1-\sqrt{m})\theta\right]^2-m}}
\end{eqnarray} 
The difference in time between $\tilde{T}^{HL}$ and $\tilde{T}^{HL}$ yields
\begin{eqnarray*}
\tilde{T}_d^{HL} &=& \tilde{T}_y^{HL} - \tilde{T}_z^{HL} \nonumber\\
 &=& \frac{1}{2}\ln{\left[\frac{1}{(1-\theta)^2} \frac{\left[1-(1-\sqrt{m})\theta\right]^2-m}{1-m}\right]}\nonumber\\
 &=& \frac{1}{2}\ln{\frac{1-2\theta+2\sqrt{m}\theta+\theta^2-2\sqrt{m}\theta^2+m\theta^2-m}{1-m-2\theta+2m\theta+\theta^2-m\theta^2}}
\end{eqnarray*}
Analogously, as in the case of switching on, one can determine whether the time difference is positive by verifying if the fraction in the logarithm is higher than or equal to $1$, which reduces to
\begin{equation*}
\theta\sqrt{m}(1-\theta)(1-\sqrt{m})\geq 0
\end{equation*}
This means that $\tilde{T}_d^{HL}$ is always positive and the time to switch off an auto-repressed gene is at most equal to the time to switch off the simple gene. Figure \ref{fig:timeNarSimpleDifference-off} confirms these results. 

In the case of vanishing leak rates $m = L = L_n=0$, the time difference between the two systems becomes zero
\begin{equation}
\tilde{T}_d^{HL} = \frac{1}{2}\ln{\frac{(1-\theta)^2}{(1-\theta)^2}} = 0
\end{equation}
Thus, for vanishing leak rates, the auto-repressed gene turns on faster compared with the simple gene, but has an equal speed when turning off. Vanishing leak rates are optimal in terms of noise and require $f(x_L)$ to be zero. This is usually difficult to achieve. For repressor genes, the gene can be turn off completely if either the Hill coefficient or  $x_L$ have high values, which comes at a high metabolic cost \citep{zabet_2009}. Even for activator genes, having a gene completely turned off can be very difficult to achieve, i.e., the regulator molecule would need to be totally absent and the affinity of RNAp for the non activated promoter should be zero. 
Putting all together, in the case of non-vanishing leak rates (sub-optimal in terms of noise), the negative auto-regulation speeds the switching  in both directions (on and off).

\subsection{Non-Instantaneous Change of Input}
\begin{figure}[htp]
  \begin{center}
	\subfigure[Fast non-instantaneous input change]{\includegraphics[scale=0.82]{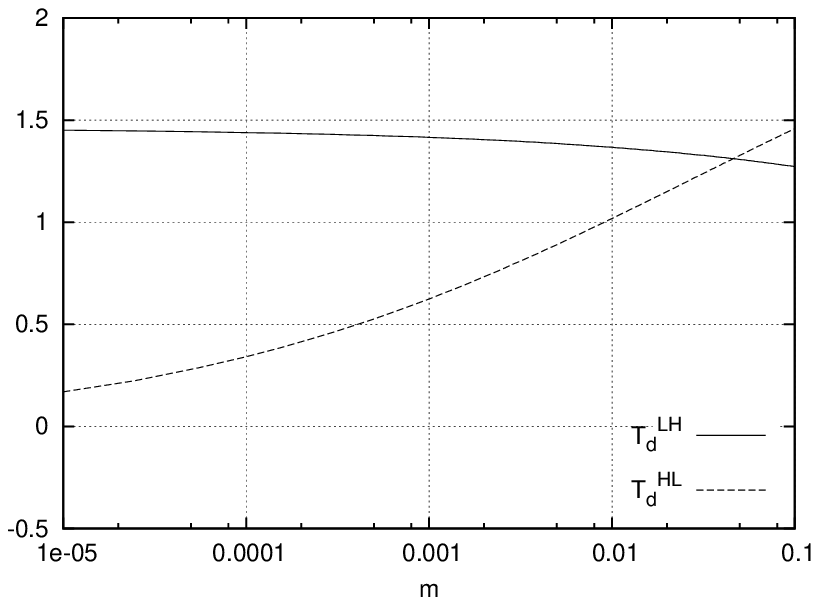}\label{fig:timeNarSimpleDifferenceNonInstant-fast}}
	\subfigure[Slow non-instantaneous input change]{\includegraphics[scale=0.82]{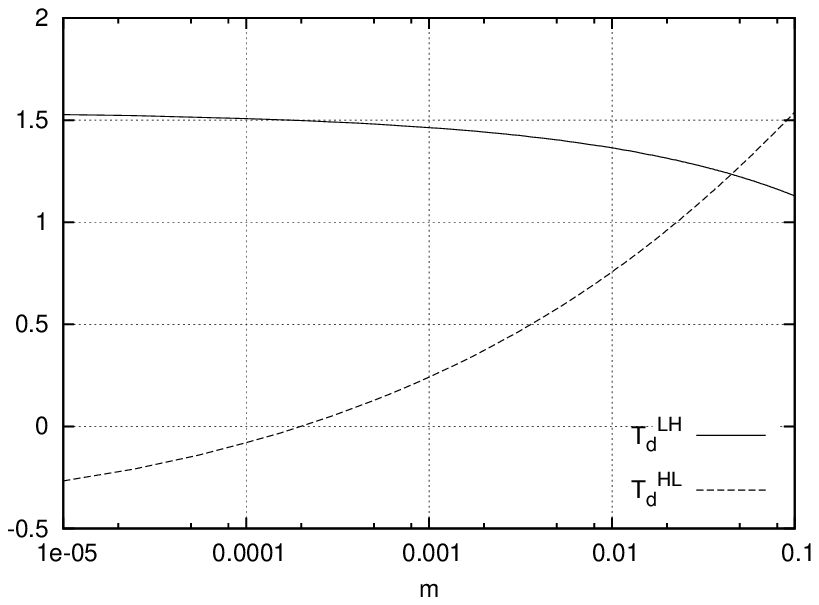}\label{fig:timeNarSimpleDifferenceNonInstant-slow}}
  \end{center}
\caption[Non-instantaneous input change and speed]{\emph{Non-instantaneous input change and speed}. The following set of parameters have been used: $\theta = 0.9$, $l=2$ and $x_H = 0.9\ \mu M$. $x_L$ was varied in the interval $x_L\in[0.0,0.2]$ and, thus, $m$ varied accordingly. The regulation threshold is selected so that the threshold resides midway between the two states of the input $K = 0.5\cdot(x_H+x_L)$ and the synthesis rate so that the cost remains fixed to $\zeta = 1.2\ \mu M min^{-1}$. Both graphs consider the activation case. \subref{fig:timeNarSimpleDifferenceNonInstant-fast} The input is assumed to change ten times faster than the output, i.e., $\mu=1\ min^{-1}$ and $\mu_x=10\ min^{-1}$. \subref{fig:timeNarSimpleDifferenceNonInstant-slow} The input is assumed to change at the same speed as the output, i.e., $\mu=1\ min^{-1}$ and $\mu_x=1\ min^{-1}$}\label{fig:timeNarSimpleDifferenceNonInstant}
\end{figure}

The assumption that $x$ changes instantaneously between $x_L$ and $x_H$ will be relaxed now. In the case of non-instantaneous input change, the solution of the differential equations \eqref{eq:modelSimpleDiff} and \eqref{eq:modelNarDiff} can only be computed numerically. For very fast, but non-instantaneous input change, one expects the behaviour to be similar to the one predicted by the instantaneous input change. Figure \ref{fig:timeNarSimpleDifferenceNonInstant-fast} confirms that the difference between the switching off time of the simple gene and the one of the auto-repressed gene is always positive. 

Usually, it is assumed that the input and the output are affected by the same decay rate (dilution) and, thus, they change at a similar speed. The numerical analysis reveals that for most of the parameter space the relationship between switching times (the signs of $T_d^{LH}$ and $T_d^{HL}$) is conserved (see Figure \ref{fig:timeNarSimpleDifferenceNonInstant-slow}). However, for no or very small leak rate ($m\leq 0.0002$ in Figure \ref{fig:timeNarSimpleDifferenceNonInstant-slow}) the switching off time seems to be increased by negative auto-regulation. This suggests that negative auto-regulation is beneficial for speed but only for non-vanishing leak rates of the output gene.

\section{Noise}
Gene expression is affected by noise \citep{spudich_1976,arkin_1998,elowitz_2002}. This noise is a consequence of the fact that genes have low copy numbers and that they are slowly expressed \citep{kaern_2005}. In the context of genes as computational units, this output noise is undesirable because it makes  difficult the assessment of the output of the gene as either low or high. 

At steady state one can compute the variance of the output species $y$ of the simple gene as \citep{paulsson_2004,shibata_2005,kampen_2007}:
\begin{equation}
\sigma_{y}^2 = y+ \left[\beta f'(x)\tau\right]^2\frac{1}{1 + \tau/\tau_x}x
\end{equation}

Usually, gene expression is modelled as a three steps process (regulation, transcription and translation), but here it is modelled as a one step process. Consequently, the noise produced by regulation and translation steps is ignored and the output noise stems mainly from transcription. Nevertheless, this assumption is often valid as shown both theoretically and experimentally \citep{even_2006,newman_2006}.

As proposed previously \citep{zabet_2009,zabet_2010b,chu_2011}, the noise will be measured by computing the variance in the high state, $y=H$, normalized by the square of the signal strength (the difference between the high and the low state), $\eta_y = \sigma_y^2/(H-L)^2 = \sigma_y^2/[H^2(1-m)^2]$.
\begin{eqnarray}
\eta_{y} &=&  \overbrace{\frac{1}{H(1-m)^2}}^{\eta_y^{\textrm{in}}} + \overbrace{\left[\frac{f'(x_H)}{f(x_H)(1-m)}\right]^2\frac{1}{1 + \tau/\tau_x}x_H}^{\eta_y^{\textrm{up}}} \nonumber\\
 & = & \frac{1}{(1-m)^2}\left[ \frac{1}{H}+ \left(\frac{f'(x_H)}{f(x_H)}\right)^2\frac{1}{1 + \tau/\tau_x}x_H \right] \label{eq:noiseSimple}
\end{eqnarray}
The noise consists of two components: the intrinsic noise, $\eta^{\textrm{in}}$, (generated by the randomness of the birth/death process) and the upstream (or extrinsic) noise, $\eta^{\textrm{up}}$, (propagated from the upstream component) \citep{elowitz_2002,swain_2002,pedraza_2005}.

In the case of the auto-repressed gene, the steady state variance of the species $z$ yields \citep{paulsson_2004,shibata_2005,kampen_2007}
\begin{equation*}
\sigma_{z}^2 = \frac{z}{1+\tau \beta f(x)H/z^2}  +  \left[\frac{\tau\beta f'(x)H/z}{\tau\beta f(x)H/z^2+1}\right]^2\frac{1}{1 + \frac{\tau}{\tau_x}\frac{1}{1+\tau \beta f(x)H/z^2}}x
\end{equation*}
Then, the noise of the auto-repressed gene (the variance in the high state normalised by the square of the signal strength) becomes
\begin{eqnarray}
\eta_{z} &=&  \overbrace{\frac{1}{H2(1-\sqrt{m})^2}}^{\eta_z^{\textrm{in}}} + \overbrace{\left[\frac{f'(x_H)}{f(x_H)\sqrt{2}(1-\sqrt{m})^2}\right]^2 \frac{1}{2+\tau/\tau_x}x_H}^{\eta_z^{\textrm{up}}}  \nonumber\\ 
 & = &  \frac{1}{2(1-\sqrt{m})^2}\left[ \frac{1}{H} + \left(\frac{f'(x_H)}{f(x_H)}\right)^2 \frac{1}{2+\tau/\tau_x}x_H \right]\label{eq:noiseNar}
\end{eqnarray}
where the following two steady state equations were used: $\beta f(x_H)\tau/H = 1$ and $\tau \beta/H=f(x_H)$.

The reliability of the analytical method for noise was validated previously by a series of stochastic simulations. The results confirmed that LNA works  well for both the simple gene \citep{zabet_2009} and the auto-repressed one \citep{thattai_2001,hayot_2004,stekel_2008}. However, for extremely strong auto-repression values ($K_n< 10^{-4}\ \mu M$), the analytical method underestimates the simulation results \citep{stekel_2008}. In this context, it is worthwhile noting that \cite{zhang_2009b} showed that, although the LNA underestimates the actual value of the noise in the auto-repressed gene, it still captures the dependence of noise on the parameters of the system.

The noise of the two systems (the simple and the auto-repressed gene) is compared by analysing the ratio for each component of the noise (intrinsic and upstream). 
\begin{equation}
\eta_c^{\textrm{in}} = \frac{\eta_{z}^{\textrm{in}}}{\eta_y^{\textrm{in}}} = \frac{(1+\sqrt{m})^2}{2}\quad \textrm{and}\quad \eta_c^{\textrm{up}} =  \frac{\eta_{z}^{\textrm{up}}}{\eta_y^{\textrm{up}}} = \frac{(1+\sqrt{m})^2}{2}\frac{1+\tau/\tau_x}{2+\tau/\tau_x} 
\end{equation}
Examining the first equation, one can notice that, for low leak rates ($m\leq 0.17$), the intrinsic component of the noise is not amplified by negative feedback.
\begin{equation*}
\eta_c^{\textrm{in}} = \frac{(1+\sqrt{m})^2}{2} \leq 1\ \Rightarrow\ m\leq (\sqrt{2}-1)^2 \approx 0.17
\end{equation*}

Furthermore, the fraction $(1+\tau_y/\tau_x)/(2+\tau_y/\tau_x)$ can never be higher than one. This yields 
\begin{equation*}
\eta_c^{\textrm{up}} \leq \frac{(1+\sqrt{m})^2}{2} \Rightarrow m\lesssim 0.17
\end{equation*}
Overall, for low leak rates, the intrinsic and the upstream components of the noise are reduced. 

\begin{figure}[htp]
  \begin{center}
	\includegraphics[scale=1.0]{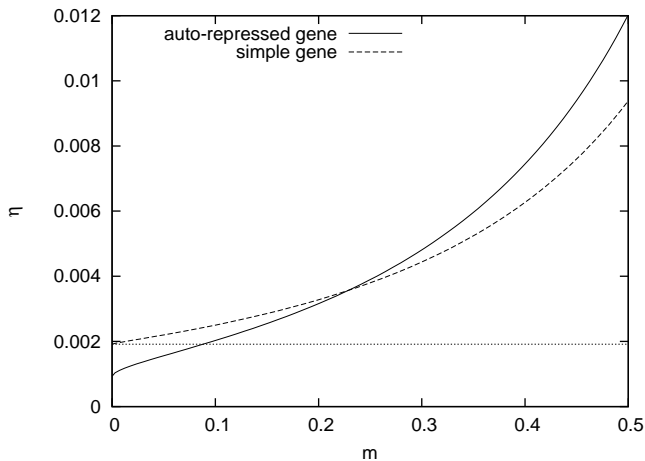}
  \end{center}
\caption[The leak rate influences the noise levels of negative auto-regulation]{\emph{The leak rate influences the noise levels of negative auto-regulation}. Increasing the leak rate reduces the accuracy. The following set of parameters was used:  $\theta=0.9$, $\mu=1 min^{-1}$, $x_H=0.9\ \mu M$, $K_n=0.01\ \mu M$ and $V=8\cdot 10^{-16}\ l$. $\beta$ was selected so that the cost remains fixed to $\zeta=1.2 \ \mu M min^{-1}$ and $K$ so that the threshold resides at the midpoint between the low and the high state of the input, $K = 0.5\cdot(x_H+x_L)$. The noise generated by $x$ is assumed to be Poissonian. The gene is activated by a regulatory input $x$. }\label{fig:noiseComparison}
\end{figure}

Figure \ref{fig:noiseComparison} confirms that, for low levels of leak rate, negative feedback enhances accuracy. This in conjunction with the fact that negative feedback increases speed for high leak rates suggest that there is a trade-off between speed and accuracy, in the sense that for lower leak rates the auto-repressed gene is more accurate than the simple gene while for higher leak rates the auto-repressed gene becomes faster. Nevertheless, these results indicate that there is a range of values for the leak rate for which the auto-repressed gene is both faster and more accurate compared to the simple one. 

A gene with negative feedback and low but non zero leak rate will display less noise ($m<0.1$ in Figure \ref{fig:noiseComparison}) and shorter response time ($m>0$ in Figure \ref{fig:timeNarSimpleDifference}) compared to the optimal configuration of the simple gene with an equivalent metabolic cost. This shows that negative feedback is able to improve the performance of a gene beyond the one of the zero leak rate configuration and without increasing the metabolic cost.


\section{Speed-Accuracy Trade-offs}

\begin{figure}[htp]
  \begin{center}
	\subfigure[low leak rate]{\includegraphics[scale=1.0]{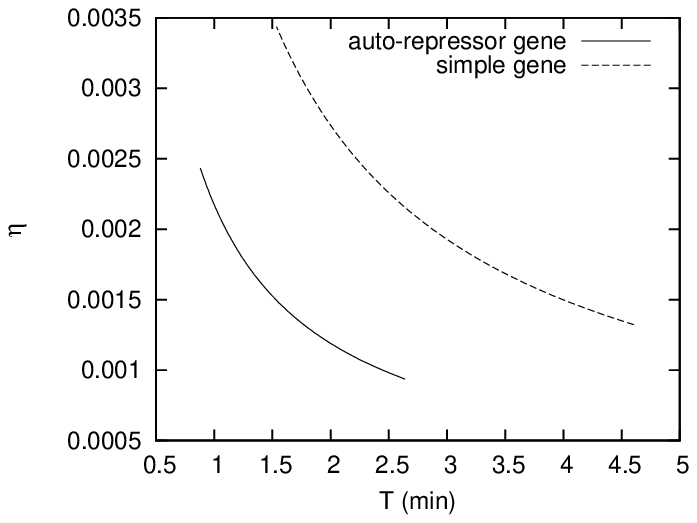}\label{fig:speedAccuracyInstant-05}}
	\subfigure[high leak rate]{\includegraphics[scale=1.0]{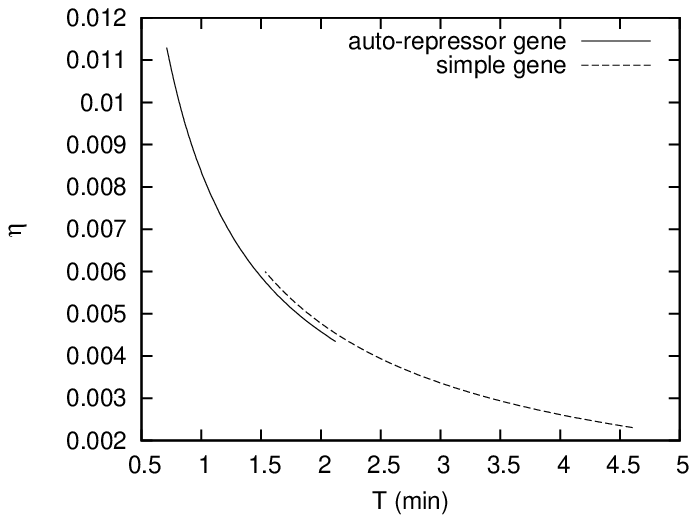}\label{fig:speedAccuracyInstant-30}}
  \end{center}
\caption[Speed-accuracy trade-off curves for instantaneous change of input]{\emph{Speed-accuracy trade-off curves for instantaneous change of input}. The following set of parameters was used:  $\theta=0.9$, $x_H=0.9\ \mu M$, $K=0.7\ \mu M$, $K_n=0.01\ \mu M$ and $V=8\cdot 10^{-16}\ l$.  $\beta$ was selected so that the cost remains fixed to $\zeta=1.2 \mu M min^{-1}$ and the decay rate $\mu$ was varied in the interval $\mu \in [0.5, 1.5]\ min^{-1}$. $G_y$ and $G_z$ are activator genes. Similar results can be obtained when the genes are repressed by the regulatory input $x$. The noise generated by $x$ is assumed to be Poissonian. Two cases were considered: \subref{fig:speedAccuracyInstant-05} $x_L=0.05\ \mu M$ and \subref{fig:speedAccuracyInstant-30} $x_L=0.30\ \mu M$}\label{fig:speedAccuracyInstant}
\end{figure}

Recently, \cite{zabet_2009} showed that the processing speed and accuracy of genes are connected, in the sense that there is a trade-off between speed and accuracy, which is controlled by the decay rate. Figure \ref{fig:speedAccuracyInstant} confirms the existence of this trade-off also in the case of the auto-repressed gene. Furthermore, the graph confirms that for low leak rates the auto-repressed gene displays a better trade-off curve compared to the simple one; see Figure \ref{fig:speedAccuracyInstant-05}. Note that not only the metabolic cost is fixed along the curve, but also the two curves display equal costs. In the case of high leak rate, the auto-repressed gene is still faster than the simple one, but this time the noise in the former is higher than the one in the latter. 

Figure \ref{fig:speedAccuracyInstant} shows that for low (but non-zero) leak rate it is always better for the gene to display negative feedback, i.e., the auto-repressed gene is both faster and more accurate than the simple one. However, for high leak rates, the cell has to choose between being fast or more accurate, by selecting whether it has negative feedback or not. This result reveals a new speed-accuracy trade-off, which is not controlled by a parameter of the system (as previously identified), but rather by the architecture of the system.

\begin{figure}[htp]
  \begin{center}
	\subfigure[instantaneous change of input]{\includegraphics[scale=1.0]{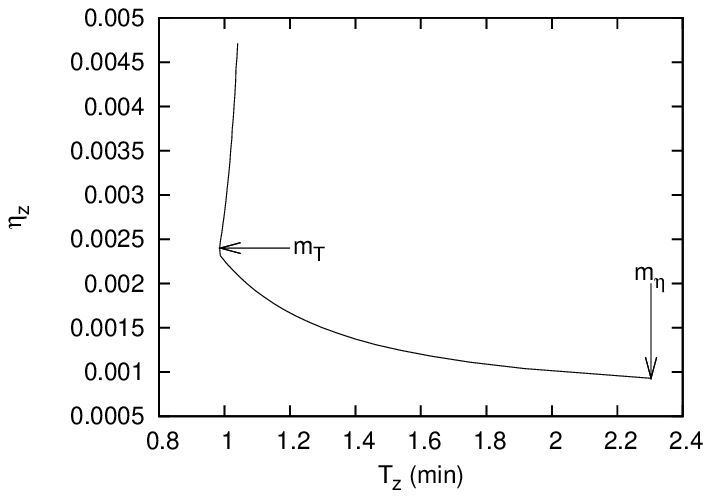}\label{fig:speedAccuracyLeakRate-instant}}
	\subfigure[exponential change of input]{\includegraphics[scale=1.0]{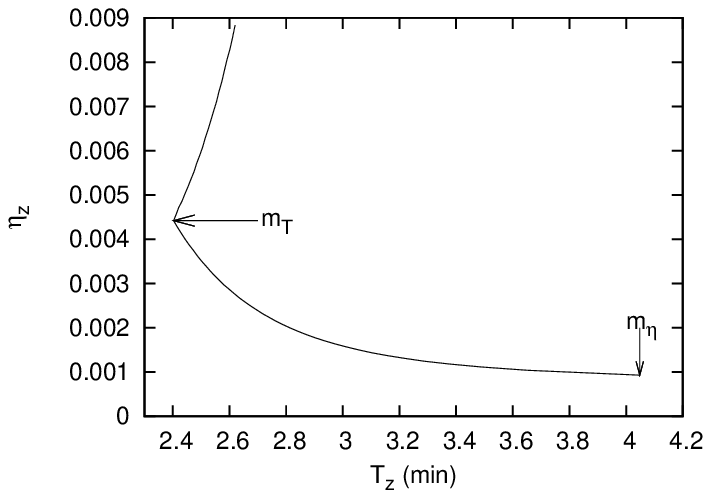}\label{fig:speedAccuracyLeakRate-exponential}}
  \end{center}
\caption[Speed-accuracy trade-off curves and leak rate]{\emph{Speed-accuracy trade-off curves and leak rate}. The following set of parameters was used:  $\theta=0.9$, $\mu =1\ min^{-1}$, $\mu_x=1\ min^{-1}$, $x_H=0.9\ \mu M$, $K=0.7\ \mu M$, $K_n=0.01\ \mu M$ and $V=8\cdot 10^{-16}\ l$. The noise generated by $x$ is assumed to be Poissonian. $\beta$ was selected so that the cost remains fixed to $\zeta=1.2 \mu M min^{-1}$ and the leak rate $m$ was varied in the interval $m \in [0.0, 0.5]$. $G_y$ and $G_z$ are activator genes.}\label{fig:speedAccuracyLeakRate}
\end{figure}

For zero leak rate, both auto-repressed and simple genes display the lowest noise configurations and this is denoted by $m=m_\eta$. It is worthwhile noting that, by increasing the leak rate, the speed gain is increased while the accuracy gain is reduced; see Figure \ref{fig:speedAccuracyLeakRate}. However, the increase in speed is achieved up to a certain maximum speed ($m_T$), above which the speed cannot be increased further by negative feedback. This indicates that there is no advantage in having leak rates higher than this optimal point in speed, while for any leak rate in the interval $m \in[m_T,m_{\eta}]$ there is a speed-accuracy trade-off. 

\section{Discussion}


Negative auto-regulation was suggested as an alternative approach to reduce the response time of a single gene. \cite{rosenfeld_2002} showed theoretically that negative auto-regulation can speed up \textit{only} the turn on response time, i.e., the turn on time of a negatively auto-regulated gene is five times smaller than the one of a simple gene. In the context of genes as computational units, the direction of switching is not important in the sense that the system needs to turn both on and off as fast as possible. One of the assumptions of \cite{rosenfeld_2002} is that the auto-repressed gene does not have a leaky expression, which is not always biologically plausible. In this contribution, it is shown that, in the case of leaky gene expression, auto-repressed genes are always faster compared to the simple ones in both turning on and off; see Figures \ref{fig:timeNarSimpleDifference} and \ref{fig:timeNarSimpleDifferenceNonInstant}.

Furthermore, several theoretical studies showed that negative auto-regulation reduces the output noise \citep{thattai_2001,paulsson_2004,shahrezaei_2008,hornung_2008,bruggeman_2009}. These results were obtained without comparing two systems with equal cost and, in fact, the negatively auto-regulated system displays a lower metabolic cost. Nevertheless, several studies \citep{paulsson_2000b,stekel_2008,zhang_2009b} compared the noise in the simple and auto-repressed gene under the assumption of fixed average number of molecules at steady state, which, in their settings, could be thought as the metabolic cost. 
A better measure of metabolic cost is the energy consumption per time unit and the scaling properties of this are  well given by the protein production rate.

Recently, \cite{chu_2011} found that negative feedback can reduce output noise in a gene cascade under the assumption of fixed production rate. However, the study of \cite{chu_2011} did not identify what parameter controls whether the auto-repressed gene has lower or higher noise compared to the simple one.
This current paper considers the case of a single gene (not a cascade) and compares the auto-repressed gene to a simple one by keeping the metabolic cost (maximum production rate) fixed. The results revealed that the leak expression rate plays a crucial role for the output noise, in the sense that, for low leak rates, the auto-repressed gene is less noisier than the simple one, while, for high leak rates, the auto-repressed gene becomes noisier; see Figure \ref{fig:noiseComparison}. 
This result can be explained by the fact that increasing the leak expression rate reduces the signal strength and, consequently, it reduces the normalisation term of the noise, $(H-L)^2$. The reduction in the signal strength is more pronounced in the case of the auto-repressed gene than in the case of the simple gene, because the low state increases faster in the former compared to the latter. Thus, negative feedback increases the sensitivity of noise to the leak expression rate. 
Note, however, that these analytical results of noise cannot be extended automatically to very strongly auto-repressed genes ($K_n<10^{-4}\ \mu M$), where the Linear Noise Approximation seems to fail to represent accurately the noise of the system \citep{stekel_2008}.

Previously, \cite{zabet_2009} showed that the speed-accuracy properties of a gene can be enhanced by reducing the leak rate, which comes at an undesirable increase in the metabolic cost. Here, it is proved that negative feedback can further enhance the speed-accuracy properties of genes without increasing the metabolic cost, by actually exploiting the non-zero leak rate of genes; see Figures \ref{fig:timeNarSimpleDifference} and \ref{fig:noiseComparison}. 

Hence, negative feedback is able to enhance the performance of a gene and this can explain why a high proportion of the bacterial genes display this mechanism. For example, in \emph{E.Coli}, $40\%$ of the genes are auto-repressed \citep{austin_2006}. Nevertheless, one question that still needs to be addressed is why only $40\%$ of the genes are negatively auto-regulated and not a higher proportion. One possible answer to this question is that for high leak rates the negative feedback only enhances speed, while at the same time reduces accuracy; see Figure \ref{fig:speedAccuracyInstant-30}. Thus, a system that displays higher leak rates has to choose between being faster or being more accurate, in the sense that the simple gene is the most accurate system, while the auto-repressed one is the fastest. It might be the case that, sometimes, it is more important for the system to be more accurate that faster, while other times to be faster than more accurate. 

Finally, the results revealed that the leak rate controls a speed-accuracy trade-off in a negatively auto-regulated gene. Particularly, there are two specific leak rate values: one which optimises the system in terms of accuracy ($m_\eta=0$) and another which optimises the system in speed ($m_T > 0$); see Figure \ref{fig:speedAccuracyLeakRate}. Increasing the leak rate above $m_T$ results in a system that is not faster than the optimal system in speed, but which has higher noise levels. This means that the leak rate determines two speed-accuracy trade-off curves, namely: \one $m\in[0,m_T]$ where the system can be optimised in speed or accuracy and \two $m\in[m_T,1]$ where the system is always suboptimal in terms of noise and speed compared to the first case. Hence, the optimal parameter set for an auto-repressed gene consists of leak rates smaller than or equal to $m_T$, while for higher leak rates the auto-repressed gene becomes suboptimal.

To sum up, negative feedback further introduces additional speed-accuracy trade-offs for genes as information processors. These trade-offs manifest through either sets of parameters (such as leak rate) or the actual architecture of the system (as either with or without feedback). Various configuration of parameters and architectures that are encountered in today's living organisms are results of long time evolution towards an optimal design.  




\section{Acknowledgements}
The author would like to thank Dr Dominique F.Chu, Professor Andrew N.W. Hone, Dr Boris Adryan and the anonymus reviewers for useful comments which lead to improvements of the manuscript and Felicia Dana Zabet for proofreading the manuscript.  

\bibliographystyle{elsarticle-harv}
\bibliography{nar_paper_jtb_rev2.bib}

\end{document}